\documentclass[aps,prl,twocolumn,showpacs,floatfix,showkeys]{revtex4-1}

\usepackage{amsmath}
\usepackage{graphicx}
\usepackage{color}
\usepackage{hyperref}
\usepackage{float}
\usepackage{amssymb}

\begin{document}

\setlength{\abovedisplayskip}{3pt}
\setlength{\belowdisplayskip}{3pt}

\title{Negative excitonic diffusion in transition metal dichalcogenides}

\author{Roberto Rosati}

\author{Ra\"ul Perea-Caus\'in}

\author{Samuel Brem}

\author{Ermin Malic}

\affiliation{Chalmers University of Technology, Department of Physics,
412 96 Gothenburg, Sweden}

\begin{abstract}
While exciton relaxation in transition metal dichalcogenides (TMDs)  has been intensively studied, spatial exciton propagation has received only little attention - 
in spite of being a key process for optoelectronics and having already shown interesting unconventional behaviours (e.g. spatial halos).
Here, we study the spatiotemporal dynamics in TMDs  and track the way of optically excited excitons in time, momentum, and space. In particular, we investigate the
temperature-dependent exciton diffusion including the remarkable exciton landscape constituted by bright and dark states. 
Based on a fully quantum mechanical approach, we show at low temperatures an unexpected \textit{negative transient diffusion}. 
This phenomenon can be traced back to the existence of dark exciton states in TMDs and is a result of an interplay between spatial exciton diffusion and intervalley exciton-phonon scattering. 
\end{abstract}

\maketitle

Transition metal dichalcogenides (TMDs) have attracted much attention \cite{Mueller18,Wang18,Xiao12,Chernikov14,He14,Yu15,Steinhoff16,Wang17,Deilmann17,Feierabend17,Selig18,Brem18,Niehues18,Deilmann19,Merkl19} in 
particular due to their remarkable exciton landscape 
including bright as well as momentum- and spin-dark exciton states \cite{Mueller18,Yu15,Wang17,Deilmann19}. 
Recently, their spatiotemporal exciton dynamics has been studied \cite{Kumar14,Mouri14,He15,Kato16,Yuan17,Cadiz17,Kulig18}, showing peculiar effects including a distinct diffusion of bright and spin-dark excitons at low temperatures \cite{Cadiz17} as well as
the formation of spatial rings at higher excitation densities \cite{Kulig18}. The origin of the latter has been suggested to stem from phonon winds \cite{Glazov19} or 
efficient thermal drifts \cite{Perea19}.

\begin{figure}[t!]
\centering
\includegraphics[width=\linewidth]{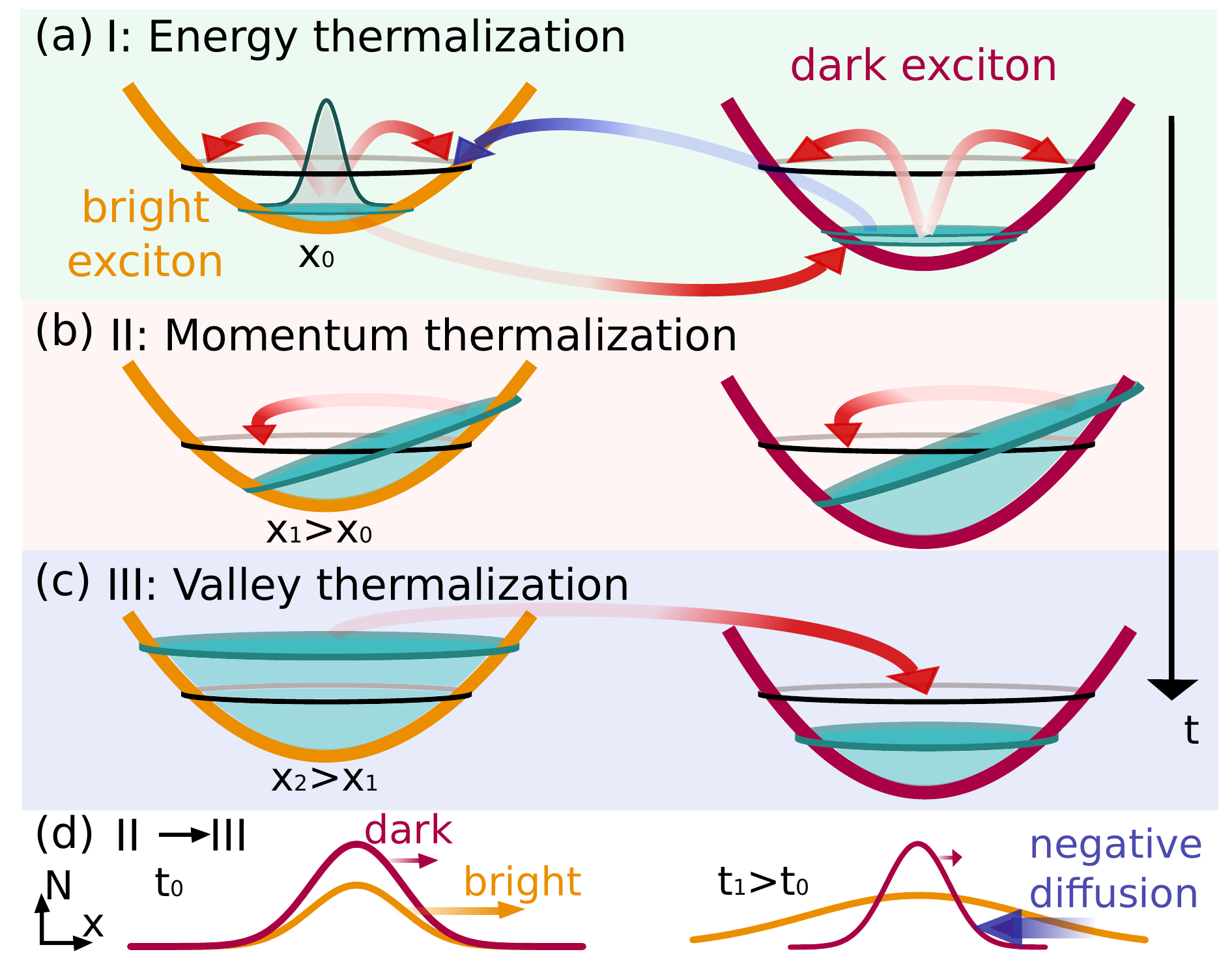}
\caption{Schematic illustration of spatiotemporal exciton dynamics in TMDs. Exciton diffusion results in anisotropic distributions at different space points 
as well as in non-equilibrium distributions between bright and dark exciton states. As a result, the dynamics is characterized by three temporally subsequent phases describing 
(a) energy, (b) momentum, and (c) valley thermalization. 
The black circles visualize the equilibrium distribution and arrows indicate exciton-phonon scattering channels. 
(d) Different exciton diffusion velocities can require back-diffusion processes (blue arrow) to establish spatial equilibrium distributions.
\label{fig:Fig1}}
\end{figure}

In this work, we shed light on the impact of the versatile exciton landscape including bright and momentum-dark states on exciton diffusion in TMDs in different temperature regimes. 
After optical excitation, excitons thermalize in energy on a quick timescale of hundreds of femtoseconds at room temperature \cite{Selig18,Brem18} 
[Fig. \ref{fig:Fig1} (a)]. However, spatial diffusion  can result also on longer timescales in anisotropic exciton 
occupations, since excitons with same energy but different momentum orientation propagate towards different spatial directions - similarly to electrons 
in quantum wells \cite{Steininger96b,Knorr98}. 
Momentum thermalization at each space point drives toward an isotropic distribution [Fig. \ref{fig:Fig1}(b)]. Furthermore, 
different diffusion velocities of dark and bright excitons can result in non-equilibrium local distributions among different valleys. Here, 
a space-dependent equilibration of bright and dark states leads toward a valley-thermalized occupation [Fig. \ref{fig:Fig1}(c)]. 

Based on a fully quantum mechanical approach \cite{Selig18,Brem18,Merkl19}, this work provides microscopic insights into the interplay of exciton diffusion and thermalization. 
We resolve the evolution of optically excited, spatially localized excitons in time, momentum, 
energy, and space taking into account bright and momentum-dark excitonic states. 
We predict unexpected spatiotemporal dynamics including the emergence of transient negative diffusion, where excitons shrink their spatial densities, apparently moving back towards
the excitation center [cf. blue arrow in Fig. \ref{fig:Fig1}(d)].

\textit{Theoretical approach:} Our goal is to study on a microscopic footing the spatiotemporal dynamics of excitons in the exemplary hBN-encapsulated tungsten disulfide (WS$_2$) monolayers. 
Considering the single-particle dispersion \cite{Kormanyos15} and solving the exciton Wannier equation \cite{Haug09,Selig16,Selig18,Brem18}, we obtain a set of exciton states
$\vert\alpha\rangle\equiv\vert\mathbf{Q},v\rangle$ characterized by the valley index $v$, the center-of-mass momentum $\mathbf{Q}$
and the exciton energy $\varepsilon_\alpha=E_v+\hbar^2|\mathbf{Q}|^2/(2M_v)$ with $M_v$ as the total valley-dependent mass. Due to considerable energy separations, 
we restrict our attention to the $1s$ states of the bright excitons ($KK$) and the momentum-dark excitons ($KK^\prime$, $K\Lambda$) lying approximately 51.5 and 30.5 mev below $KK$, respectively. 
Taking these states as basis, we introduce the coefficients $\rho_{\alpha \alpha^\prime}$ of the single-particle density matrix \cite{Rossi02,Kira06,Malic13} of incoherent excitons,
$\rho_{\alpha \alpha^\prime} = \rho^{vv^\prime}_{\mathbf{Q} \mathbf{Q}^\prime}=\left\langle \hat X^\dagger_{\alpha^\prime} X_{\alpha} \right\rangle $,
where $X^{(\dagger)}_{\alpha}$ are 
annihilation (creation) operators for the state $\vert\alpha\rangle$. 
Then, we introduce the excitonic intravalley Wigner function 
$N^v_{\mathbf{Q}}(\mathbf{r},t)\equiv \sum_{\mathbf{q}^\prime} \rho^{vv}_{\mathbf{Q}-\mathbf{q}^\prime/2,\mathbf{Q}+\mathbf{q}^\prime/2} 
e^{\imath\mathbf{q}^\prime\cdot\mathbf{r}}$, which summed over $\mathbf{Q}$ provides the 
intravalley spatial density 
$N_v(\mathbf{r},t)\equiv \frac{1}{V}\sum_{\mathbf{Q}} N^v_{\mathbf{Q}}(\mathbf{r},t)$. 
The total excitonic amount $n_v(t)$ 
in valley $v$ reads 
$n_v(t)=\int N_v(\mathbf{r},t) d\mathbf{r}=\sum_{\mathbf{Q}} \rho^{vv}_{\mathbf{Q}\mathbf{Q}}$ 
and is determined by the diagonal density matrix elements.
In contrast, the off-diagonal terms 
induce an explicit dependence on $\mathbf{r}$, i.e. they result in a spatial inhomogeneity. 

Now, we introduce an equation of motion
for the spatiotemporal dynamics of excitons by exploiting the Heisenberg equation and the many-particle Hamilton operator \cite{Kira06,Malic13,Haug09}. The derived semiconductor Bloch equation \cite{Selig18,Merkl19} can 
then be transformed in the Wigner representation \cite{Hess96,Jago19} and reads in the low excitation regime:
\begin{align}
\begin{split}\label{SBE}
\partial_t N^v_\mathbf{Q}(\mathbf{r},t)=& \left(\frac{\hbar \mathbf{Q}}{M_v}\cdot \nabla - \gamma \delta_{\mathbf{Q},0}\delta_{v,KK} \right)N^v_\mathbf{Q}(\mathbf{r},t)\\ 
&+\Gamma^{v;KK}_{\mathbf{Q};0} |p_{0}(\mathbf{r},t)|^2\!\!+\!\!\left.\partial_t N^v_\mathbf{Q}(\mathbf{r},t)\right|_{scat}\,.\\
\end{split}
\end{align}
The first term indicates the free evolution of excitons $\nabla_{\mathbf{Q}}\varepsilon_{\mathbf{Q},v}=\hbar \mathbf{Q}/M_v$, while the second term takes into account the losses due to the direct photoluminescence $I_{\textrm{PL}}(\mathbf{r},t)=\gamma N^{KK}_{\mathbf{Q}\approx 0}(\mathbf{r},t)$. Here,  $\gamma$ describes the radiative recombination rate within the light cone ($\delta_{\mathbf{Q},0}\delta_{v,KK}$) \cite{Selig16,Selig18,Brem18}. Effects of 
phonon-assisted radiative recombination are beyond the scope of this work \cite{Brem19}. 

The
first contribution in the second line of Eq. (\ref{SBE}) describes the formation of incoherent excitons due to phonon-driven transfer from 
the excitonic polarization $p_{\mathbf{Q}\approx 0}(\mathbf{r}, t)$ (referred to in literature as coherent excitons \cite{Selig18}). 
The latter are optically excited by an electromagnetic
field $\mathbf{A}(\mathbf{r},t)$ through
$\left.\partial_t p_{\mathbf{Q}}(\mathbf r, t)\right|_{opt}\propto \mathbf{M}\cdot\mathbf{A}(\mathbf{r},t) \delta_{\mathbf{Q}, 0}$, with $\mathbf{M}$ 
depending on optical matrix elements and excitonic wave functions \cite{Selig18,Brem18}. 
The process is driven by exciton-phonon scattering with the rates $\Gamma^{vv^\prime}_{\mathbf{Q}\mathbf{Q}^\prime}$ 
describing scattering from state $\vert \mathbf{Q}^\prime v^\prime\rangle$ to $\vert \mathbf{Q} v\rangle$  via interaction with phonons \cite{Selig18,Brem18}.
Since coherent excitons decay on an ultrafast timescale of 10-100 fs\cite{Selig18}, it is the incoherent exciton distribution that determines the diffusion. 

Finally, the last term in Eq. (\ref{SBE}) describes the scattering contribution. Here, we restrict our attention to the low-excitation regime, where the main source of scattering
is given by exciton-phonon interactions. The intra- ($v=v^\prime$) and inter-valley ($v\neq v^\prime$) contribution can be written as 
$\left.\partial_t N^v_\mathbf{Q}(\mathbf{r},t)\right|_{scat}=\sum_{v^\prime}\left.\partial_t N^v_\mathbf{Q}(\mathbf{r},t)\right|_{v^\prime}$,
where $\left.\partial_t N^v_\mathbf{Q}(\mathbf{r},t)\right|_{v^\prime}$ indicates the dynamics of $N^v$ induced by the interaction with $N^{v^\prime}$.
The corresponding equation of motion reads
\begin{equation}\label{dFvvP}
\left.\partial_t N^v_\mathbf{Q}(\mathbf{r},t)\right|_{v^\prime}\!\! = \!\!\sum_{\mathbf{Q}^\prime} \!\!\left[\Gamma^{vv^\prime}_{\mathbf{Q}\mathbf{Q}^\prime} N^{v^\prime}_{\mathbf{Q}^\prime}(\mathbf{r},t) \!-\!\Gamma^{v^\prime v}_{\mathbf{Q}^\prime\mathbf{Q}}N^v_{\mathbf{Q}}(\mathbf{r},t) \right],
\end{equation}
where the first (second) term describes the in- (out-) scattering dynamics of the Wigner function
$N^v_{\mathbf{Q}}$. Using the index shift $\mathbf{Q} \leftrightarrow \mathbf{Q}^\prime$ one can show for the exciton density
$
\left.\partial_t N_v(\mathbf{r},t)\right|_{v^\prime}\equiv \frac{1}{V}\sum_{\mathbf{Q}}\left.\partial_t N^v_\mathbf{Q}(\mathbf{r},t)\right|_{v^\prime}\!\! = \!\!-\partial_t N_{v^\prime}(\mathbf{r},t)\vert_{v}\,.
$
It follows immediately that intravalley scattering does not change $N_v(\mathbf{r},t)$, i.e. $\partial_t N_v(\mathbf{r},t)\vert_{v}=0$, 
as expected for broad densities \cite{Rosati13b,Rosati14b}. 
Although having no direct contribution, the intravalley scattering may have a considerable impact on the spatial distribution $N_v(\mathbf{r})$,
since locally (i.e. in every position $\mathbf{r}$) these scattering channels redistribute
the Wigner function $N^v_{\mathbf{Q}}$ in the momentum $\mathbf{Q}$ toward the local equilibrium  distribution 
$N^{v\circ}_{\mathbf{Q}}(\mathbf{r},t)\propto \textrm{Exp}[-\varepsilon_{\mathbf{Q}v}/(k_BT)]$.
By studying how the difference between $N^v_{\mathbf{Q}}$ and $N^{v\circ}_{\mathbf{Q}}$ evolves \cite{Hess96}, it can be shown that the dynamics of the spatial distribution
$N_v$ in the absence of intervalley scattering mechanisms is given by Fick's law
\begin{equation}\label{Fick}
\partial_t N_v(\mathbf{r},t)=D_v \Delta_{\mathbf{r}} N_v(\mathbf{r},t).
\end{equation}
Here, $D_v=1/2\langle \tau^v_{\mathbf{Q}}\hbar^2 Q^2/M_v^2 \rangle\vert^v_{\mathbf{Q}}$ is the diffusion coefficient with $\tau^v_{\mathbf{Q}}=\sum_{\mathbf{Q}^\prime} \Gamma^{vv}_{\mathbf{Q}^\prime\mathbf{Q}}$ 
providing the $\mathbf{Q}$-dependent relaxation time induced by intravalley processes. The introduced expectation value $\langle f_{\mathbf{Q}}\rangle\vert^v_\mathbf{Q}$ provides the average of $f_{\mathbf{Q}}$ assuming a (local) thermalized distribution 
$\langle f_{\mathbf{Q}}\rangle\vert^v_\mathbf{Q}=\sum_{\mathbf{Q}}f_{\mathbf{Q}}\textrm{Exp}
\big(-\frac{\varepsilon_{\mathbf{Q}v}}{k_BT}\big)/\sum_\mathbf{Q}\textrm{Exp}\big(-\frac{\varepsilon_{\mathbf{Q}v}}{k_BT}\big)$. Under the assumption of constant relaxation times 
$\tau^{vv}_{\mathbf{Q}}\approx \tau_v$ the well-known steady-state  relation $D_v=\tau_v k_B T/M_v$ can be recovered. 

The evolution
of $N_v(\mathbf{r},t)$ is symmetric in angle, preserves the location of its center, and broadens in space, cf. Eq. (\ref{Fick}). 
In order to quantify this diffusion process, we introduce a width $w_v$ of the distribution $N_v$ which is proportional to the variance, 
$w_v^2= \int \mathbf{r}^2N_v(\mathbf{r},t) d\mathbf{r}/n_v$. According to Fick's law [Eq. (\ref{Fick})], 
confined spatial distributions behave as 
$N_v(\mathbf{r},t)\propto \exp\left[-r^2/w_v^2(t)\right]$ with $w_v^2(t)=w_{v,0}^2+4 D_v t$ \cite{He15,Kulig18}, 
where $w_{v,0}^2$ is the initial width. It follows that the diffusion coefficient can be defined as $D_v=\frac{1}{4}\partial_tw_v^2$, i.e. it corresponds to the slope of the temporal evolution of the squared width $w^2_v$.
Although $D_v$ is typically a constant in the steady-state regime, it is expected to undergo a non-trivial evolution directly after the excitation. As a remarkable example, in the so-called ballistic scenario, 
where many-particle scattering is negligible, one finds that the diffusion coefficient $D_v$ is directly proportional to time $t$ \cite{Steininger96b}.
The intravalley scattering leads the system from a ballistic evolution to a conventional diffusion by
acting as a frictional mechanism counteracting the free diffusion. When the intervalley contribution 
becomes of the
same order of magnitude as the intravalley one, intriguing phenomena are expected including negative diffusion effects - as will be discussed below.

\begin{figure}[t]
\centering
\includegraphics[width=\linewidth]{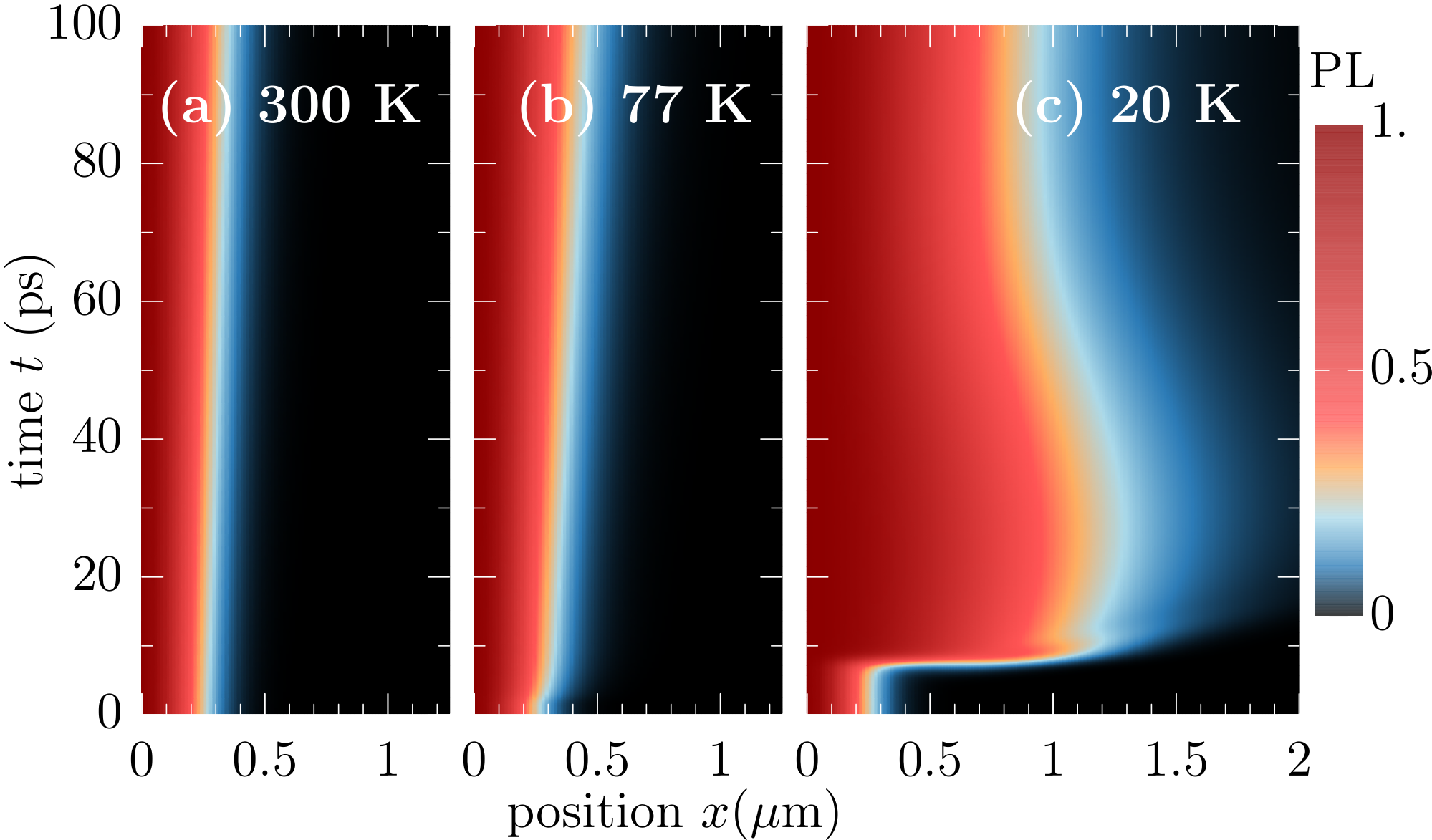}
\setlength{\abovecaptionskip}{-10pt}
\setlength{\belowcaptionskip}{-10pt}
\caption{Normalized photoluminescence in a 
hBN-embedded monolayer of WS$_2$ at (a) 300 K, (b) 77K and (c) 20 K as a function of spatial position $x$ and time $t$. 
At the lowest temperature, we observe an unexpected narrowing of the PL indicating a negative exciton diffusion.
\label{fig:Fig2}}
\end{figure}

\textit{Spatiotemporal photoluminescence:} Now, we exploit the theoretical approach described above to describe the spatiotemporal dynamics in monolayer TMDs. First, we create a non-equilibrium through 
an optical excitation which is resonant to the A exciton and centered in time (around $t_0=$0.2 ps with $\mathbf{A}(\mathbf{r},t)$ having FWHM of 100 fs) 
and space (around $\mathbf{r}$=0 with FWHM of 0.5 $\mu$m). Then, we evaluate Eqs. (\ref{SBE}) and (\ref{dFvvP}) to describe transient changes of the diffusion coefficients including phonon-assisted 
charge transfer between bright and momentum-dark excitons.
Figure \ref{fig:Fig2} shows the space- and time-resolved photoluminescence $I_{PL}$ normalized to the center of the peak. It illustrates the increase of the lateral size of the spatial exciton distribution at different temperatures. At 300 K, there is only a modest and constant broadening, while at 77K, one can see a more pronounced increase of the spatial size due to the weaker counteracting exciton-phonon scattering at low temperatures. 
Interestingly, different timescales are observable, exhibiting a much faster diffusion in the first few picoseconds followed by a slower broadening. 

Further decreasing the temperature to 20 K, we find in addition to an unexpected fast broadening at around 8 ps an even more surprising behaviour: 
After approximately 30 ps, 
the photoluminescence changes its flux, i.e. the PL goes back in space towards the center of the distribution rather than diffusing away from it. 
This is due to an efficient \textit{back-diffusion} or according to Eq. (\ref{Fick}), \textit{negative} diffusion. 
Although also observed in single-species studies \cite{Zhao03,Zhao05}, negative diffusions 
appear usually in multi-component systems (also called uphill diffusion, see e.g. \cite{Gupta70,Wolf93,Lauerer15}).
In our case, the multi-component nature is induced by the remarkable multi-valley exciton landscape displayed in TMDs. We will show below that the efficient intervalley exciton-phonon scattering
is the origin of both the back-diffusion as well as the sharp increase at about 8 ps. Note that the observed diffusion delay stems from the time required to absorb intervalley phonons 
from higher-energy dark exciton states.

\begin{figure}[t]
\centering
\includegraphics[width=\linewidth]{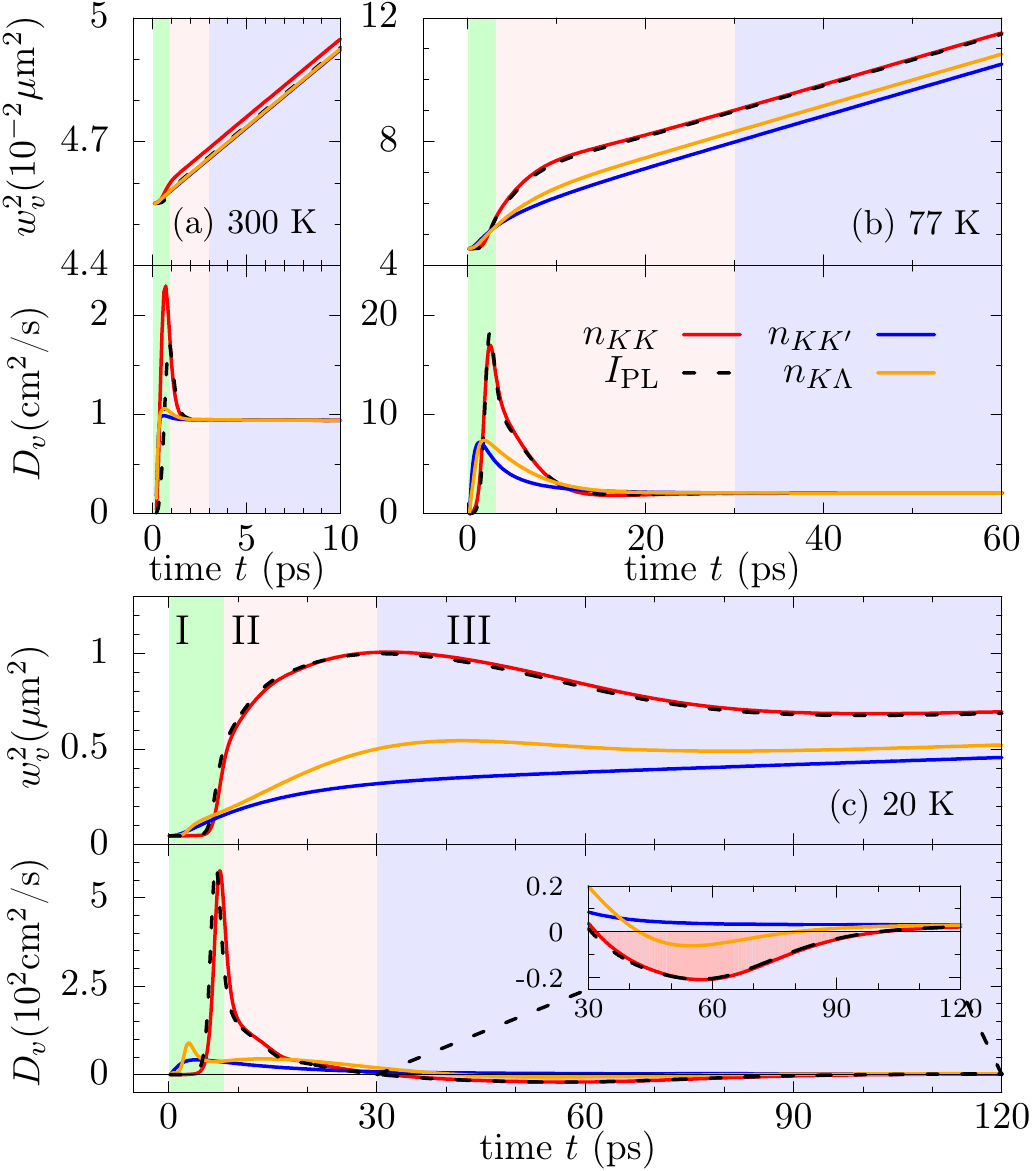}
\setlength{\abovecaptionskip}{-10pt}
\setlength{\belowcaptionskip}{-10pt}
\caption{Temporal evolution of squared spatial width $w_v^2$ of excitonic distributions and the associated transient diffusion coefficient $D_v$ 
of $KK$, $KK^\prime$ and $K\Lambda$ excitons shown at (a) 300 K, (b) 77K, and (c) 20 K. The evolution of the direct PL is illustrated with the dashed black line.
The three phases characterizing the spatiotemporal dynamics are denoted with the green, red and blue background color, respectively (as introduced in Fig.\ref{fig:Fig1}).
\label{fig:Fig3}}
\end{figure}

\textit{Valley-dependent exciton diffusion:} To understand the interesting spatiotemporal evolution of the PL, we perform now a quantitative analysis of the shape of excitonic densities including bright and dark states. 
Figure \ref{fig:Fig3} illustrates the time evolution of the squared spatial width
$w_{v}^2$ and the resulting diffusion coefficient $D_v$ for $KK$, $KK^\prime$ and $K\Lambda$ excitons. 
The PL (dashed lines) follows the dynamics of $n_{KK}$. At room temperature [Fig. \ref{fig:Fig3}(a)] we observe an almost identical behavior for all 
three types of excitons. The diffusion coefficients $D_v$ 
becomes very quickly time-independent and reaches a stationary value of about 1 cm/s$^2$. 
This is a consequence of a very efficient exciton-phonon scattering at 300 K that results in an ultrafast equilibration of all intra- and intervalley exciton states, before a 
considerable spatial separation between valleys could appear.
This is not the case anymore at 77 K [Fig. \ref{fig:Fig3}(b)], where we observe valley-dependent values. 

We find three well-distinguished phases determining the temporal evolution of exciton diffusion in TMDs:
I) an initial phase with a valley-dependent fast increase of diffusion coefficients (green-shaded background), II)
a transient phase where the differences between valleys decrease (red-shaded background), and III) a final phase, where an effective 
stationary diffusion coefficient is reached (blue-shaded background). 
It is the relatively long duration of phase II, where the diffusion coefficients are still valley-intrinsic, which gives rise to the observed different widths 
$w_v^2$ for different exciton densities. These differences are essentially preserved in phase III, i.e. the $w^2_v$ trajectories 
tend to be parallel resulting in similar diffusion coefficients $D_v$.

Decreasing the temperature to 20 K [Fig. \ref{fig:Fig3}(c)], both the maximal values of coefficients $D_v$ and the spatial separation between  
valley-dependent widths $w^2_v$ further increase, reflecting the strongly reduced efficiency of exciton-phonon scattering counteracting diffusion 
and valley redistribution of excitons. After the initial steep increase, the diffusion coefficients of both $KK$ and $K\Lambda$ excitons undergo a subsequent 
decrease eventually leading to negative values at around 30-100 ps for $KK$ excitons (see inset). 
The appearance of negative diffusion explains the features observed in the PL in Fig. \ref{fig:Fig2}.

\begin{figure}[t]
\centering
\includegraphics[width=\linewidth]{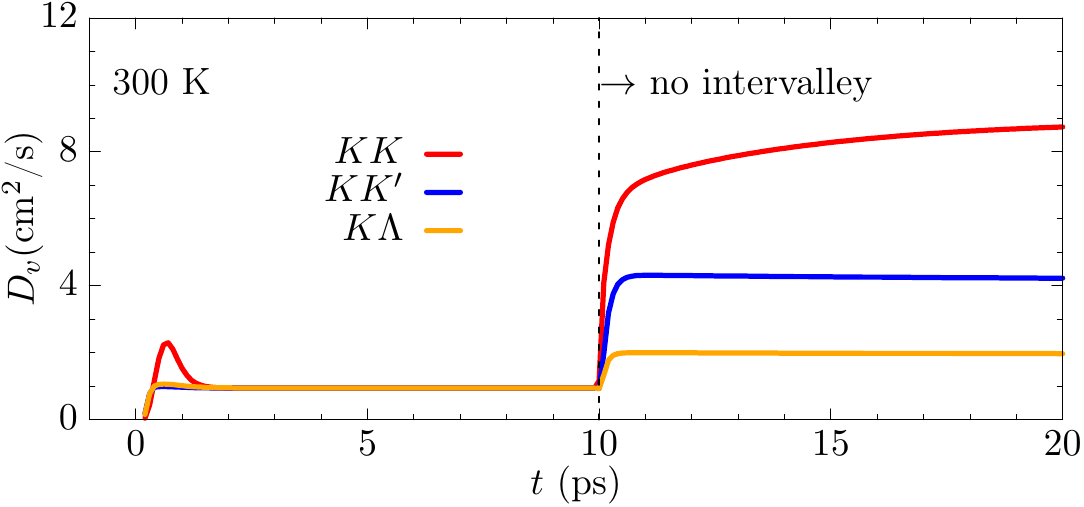}
\setlength{\abovecaptionskip}{-10pt}
\setlength{\belowcaptionskip}{-10pt}
\caption{Diffusion coefficient of $KK, KK^\prime, K\Lambda$ excitons at 300 K, where intervalley 
channels are switched off after 10 ps. 
\label{fig:Fig4}}
\end{figure}

\textit{Intervalley exciton-phonon scattering:} To better understand the mechanism underlying the predicted negative diffusion, we investigate the role of intervalley exciton-phonon scattering. Figure \ref{fig:Fig4} illustrates the full temporal evolution of $D_v$ at 300 K in the gedanken experiments, where we artificially switch off the intervalley scattering after 10 ps.
We clearly observe that without intervalley scattering one would have 
 pronounced valley-dependent diffusion coefficients $D_v$ reflecting different effective masses of the involved valleys. This already shows the crucial role of intervalley scattering for exciton diffusion. 

To be able to quantitatively understand how the intervalley scattering affects the diffusion at low temperatures,
we introduce the \textit{scattering-induced shape variation}
\begin{equation}\label{dNvvP}
\eta_v(\mathbf{r},t)=\partial_t N_v(\mathbf{r},t)\vert_{scat}-c_vN_v(\mathbf{r},t) \quad ,
\end{equation}
with the space-independent ratio $c_v=\frac{1}{n_v}\sum_{v^\prime}\int \partial_t N_v(\mathbf{r},t)\vert_{v^\prime} d\mathbf{r}$. 
The shape variation $\eta_v(\mathbf{r},t)$ is the difference between full scattering-induced density dynamics $\partial_t N_v(\mathbf{r},t)\vert_{scat}$
and 
 shape-preserving change of exciton amount. The latter only induces a variation of amplitude 
$N_v(\mathbf{r},t+\Delta t)=(1+c_{v}\Delta t)N_v(\mathbf{r},t)$ in the limit $\Delta t \to 0$.
The quantity $\eta$ alone, in contrast, would induce $N_v(\mathbf{r},t+\Delta t)=N_v(\mathbf{r},t)+\Delta t \eta_{v}(\mathbf{r},t)$. Since 
$\int d\mathbf{r}\eta_{v}(\mathbf{r},t)=0$, it follows that $\eta_v$ may describe amount-preserving variations of density $N_v$, i.e. $\eta_v$ extracts
the changes in shape of $N_v$ from $\partial_t N_v(\mathbf{r},t)\vert_{scat}$, while the variation of $n_v$ is described by $c_vN_v$. 
The shape variation $\eta_v$ describes scattering-induced spatial redistribution of density $N_v(\mathbf{r})$ from,
e.g. position $\mathbf{r}_1$ 
toward $\mathbf{r}_2$, by assuming negative (positive) values in the regions where the distribution is lost (gained).
Thus, $\eta$ directly affects the exciton diffusion process. Exploiting the general definition of the diffusion coefficient $D_v$ below Eq. (\ref{Fick}),  we introduce a \textit{scattering-induced diffusion coefficient} 
$
D^{\textrm{scat}}_{v}=\left.\frac{1}{4}\partial_t \left(\langle\mathbf{r}^2 N_v\rangle\vert^v_{\mathbf{r}}\right)\right|_{scat}\approx\frac{\int \mathbf{r}^2 \eta_v(\mathbf{r},t)d\mathbf{r}}{\int 4 N_v(\mathbf{r},t) d\mathbf{r}}\,.
$ This quantity provides
the contribution of exciton-phonon scattering to the exciton diffusion. 
Recalling that intravalley scattering does not contribute to the dynamics of $N_v$, $D^{\textrm{scat}}_{v}$ can also be seen as the direct measure for the 
intervalley scattering-induced diffusion $D^{\textrm{interv}}_{v}$. 

\begin{figure}[t]
\centering
\includegraphics[width=\linewidth]{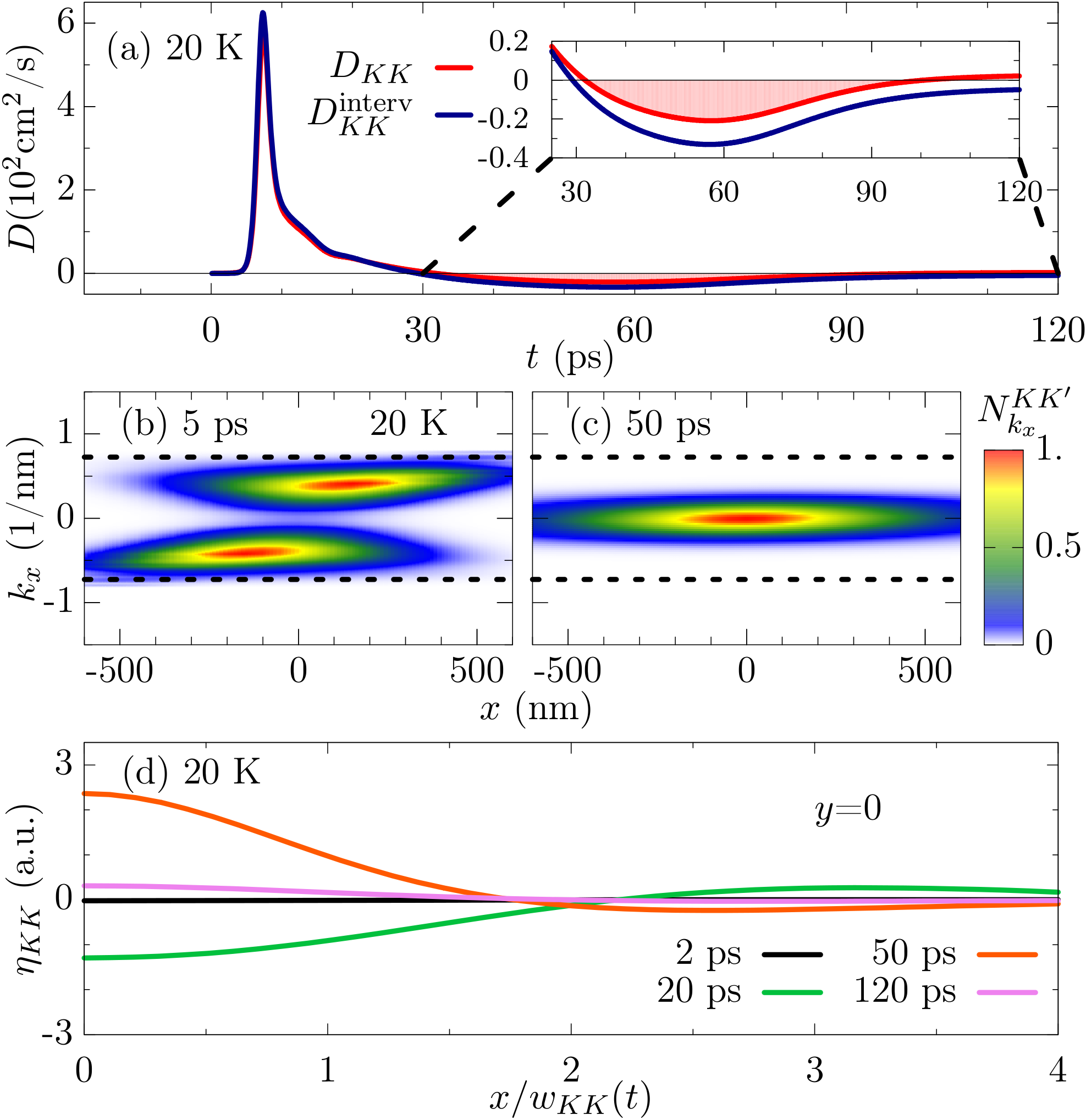}
\setlength{\abovecaptionskip}{-10pt}
\setlength{\belowcaptionskip}{-10pt}
\caption{Illustration of mechanism behind negative diffusion. 
(a) Intervalley scattering-induced diffusion coefficient
$D^{\text{interv}}_{KK}$ (see the definition in the text) in direct comparison to the full diffusion coefficient $D_{KK}$ at 20 K illustrating the crucial contribution of intervalley scattering. 
(b)-(c) Wigner distribution $N^{KK^\prime}_{k_x}(x)$ for $KK^\prime$ excitons plotted for  at 5 and 50 ps as a function of $\mathbf{k}\equiv (k_x,0)$
and $\mathbf{r}\equiv (x,0)$. The dotted line 
indicate the minimum wave-vector modulus required for intervalley process into valley $KK$ via absorption of acoustic modes.
(d) Normalized scattering-induced shape variation [cf. Eq. (\ref{dNvvP})] for $KK$ excitons
at four different times. 
\label{fig:Fig5}}
\end{figure}

Figure \ref{fig:Fig5}(a) shows a direct comparison between scattering-induced ($D^{\textrm{interv}}_{KK}$) and full diffusion coefficient $D_{KK}$ at 20 K. 
We find strong similarities between the two lines, indicating that the scattering-induced diffusion dominates the transient features of $D_{KK}$.
These include e.g. the high peak at about 8 ps, which can be explained as follows: Due to the low temperature the momentum-dark excitons
have not fully thermalized yet, having an overpopulation of states with energies high enough to scatter into $KK$ states (cf. blue arrow in Fig. \ref{fig:Fig1}(a)).

This can be observed also in Fig. \ref{fig:Fig5}(b), where the Wigner distribution $N^{KK^\prime}_{k_x}(x)$ for $KK^\prime$ excitons is plotted at 5 ps. 
Due to the reduced temperature, both energy and momentum thermalization is not finished yet.
We find a clear  anisotropy in momentum, such that there is a higher exciton occupation at positive (negative) momenta for $x>0$ ($x<0$).
The energy distribution is not thermalized around $k_x \approx 0$, but exhibits peaks at finite momenta. 

The non-equilibrium excess-energy  is induced by the polarization-to-population transfer, 
implying that the incoherent dark states formed directly after the optical excitation lay one intervalley phonon energy below the $KK$ minimum 
(i.e. the energy of coherent excitons). This creates an initial hot distribution of dark excitons, which subsequently relaxes toward the corresponding ground state. 
Note that these hot dark excitons can scatter back into  $KK$ exciton states  via absorption of intervalley phonons. As a guide to the eye, the dotted 
line in Figs. \ref{fig:Fig5}(b-c) indicates the minimum wave-vector modulus required for such an intervalley process via absorption of acoustic modes.
As shown in Fig.  \ref{fig:Fig5}(b), the distribution of these $KK^\prime$ states with high-enough energy is mostly located away from the initial
excitation spot, while the other $KK^\prime$ with smaller energies are distributed closer to $\mathbf{r}\approx 0$. 
Once these hot states scatter into valley $KK$, the density of the latter gets populated in positions which are not only bigger than $w_{KK}$ (before these intervalley
processes) but also bigger than $w_{KK^\prime}$, which is dominated by the states unable to scatter into $KK$, i.e. by those lying closer to $\mathbf{r}\approx 0$. 
As a result, $w^2_{KK}$ quickly increases
toward values higher than those of  $w^2_{KK^\prime}$ [see Fig. \ref{fig:Fig3}(c)].
The absorption of phonons with finite energy is subject to a certain delay time at small temperatures, which explains the initial slow diffusion of $KK$ excitons. 
Once occupied mostly
via scattering from high-energy dark states, the $KK$ excitons thermalize. It is during this valley-thermalization phase [cf. also Fig. \ref{fig:Fig1}(c-d)] 
that the negative transient diffusion in $D_{KK}$ appears. 
When the diffusion of $KK$ excitons shows the highest negative values, the distribution of $KK^\prime$ excitons is already quasi-thermalized both in energy and momentum, 
cf. Fig. \ref{fig:Fig5}(c).

The driving force behind the remarkable negative diffusion are intervalley exciton-phonon scattering processes, as the similarity of $D^{\textrm{interv}}$ 
with the scattering-induced diffusion $D$ shows if Fig. \ref{fig:Fig5}(a) (see also the inset).
Note that the effect of $D^{\textrm{interv}}_v$ becomes smaller for excitons with a higher population, i.e., the energetically lowest and 
thus the highest occupied $KK^\prime$ excitons
have a negligible $D^{\textrm{interv}}$ (not shown), resulting in a smoother evolution of $w^2_{KK^\prime}$ and $D_{KK^\prime}$ [Fig. \ref{fig:Fig3}(c)]. 
In fact, it can be shown that 
the mutual interaction between two valleys $v_1$ and $v_2$ leads to
$\eta_{v_1}\approx-\eta_{v_2}$. Inserting the last approximation in the $\eta$-dependent definition of $D^{\textrm{interv}}$, 
one finds $|D^{\textrm{interv}}_{v_1}|=|D^{\textrm{interv}}_{v_2}|n_{v_2}/n_{v_1}$.
As a consequence, for $n_{v_1} \ll n_{v_2}$ it follows that $|D^{\textrm{interv}}_{v_1}|\gg|D^{\textrm{interv}}_{v_2}|$.
This explains why the negative diffusion region only appears for lower-populated exciton valleys [Fig. \ref{fig:Fig3}(c)].

The different temporal behaviour displayed by $D^{\textrm{interv}}_{KK}$ is induced by different qualitative trends of the shape variation $\eta_{KK}$, 
cf. Fig. \ref{fig:Fig5}(d). Here, we have normalized $\eta_{KK}$ with respect to $n_{KK}(0,t)$ and considered the dimension-free position $x/w_{KK}(t)$ to compare spatial distributions with different
height and width. At 2 ps, the timescale is too short for the absorption of phonons, hence both $D^{\textrm{interv}}$ and $\eta$ are negligible.
In contrast, at 20 ps
$D^{\textrm{interv}}$ is very high. Accordingly, $\eta$ shows the conventional diffusive shape with a transfer of density from the center [$\eta(x)<0$ for $x\approx 0$] toward the
tails ($\eta>0$ for $x\gtrsim 2w_{KK}$). However, at 50 ps there is a remarkable change of sign in $\eta$, which is now positive in the center and negative in the tail. 
This implies 
an uphill transfer of density from the tails toward the center, i.e. a negative diffusion, which is in fact displayed in Fig. \ref{fig:Fig5}(d) at 50 ps. 
At 120 ps, there is a similar behaviour of $\eta$, however with a reduced magnitude. This implies a $D^{\textrm{interv}}$ with a reduced negativity (in agreement with Fig. \ref{fig:Fig5}(a)) that is  overcompensated by  
other mechanisms resulting in an overall positive diffusion, cf. the red line in Fig.  \ref{fig:Fig5}(a).

\textit{Conclusions:} We have shown that the spatiotemporal exciton dynamics in transition metal dichalcogenides can result in an unexpected negative exciton diffusion, 
i.e. a shrinking  of the spatial exciton density. 
Based on a fully quantum mechanical approach providing microscopic insights into time-, momentum- and space-resolved exciton dynamics, 
we ascribe this behaviour to the interplay of valley-intrinsic diffusion and intervalley thermalization processes. The key ingredient is the 
remarkable excitonic landscape of TMDs containing bright and lower-lying dark states. Our work sheds light on the emerging field of spatiotemporal dynamics in 
atomically thin materials and may trigger new experimental and theoretical studies on valley-dependent exciton diffusion.


\section*{Acknowledgements}
This project has received funding from the Swedish Research
Council (VR, project number 2018-00734) and the European Union's Horizon 2020
research and innovation programme under grant agreement
No 785219 (Graphene Flagship).

\bibliography{nanoscaleBib}

\end{document}